\documentclass[twocolumn,usenames,dvipsnames]{aastex631}
\usepackage{newtxtext,newtxmath}

\usepackage{amsmath, booktabs, multirow, rotating}
\usepackage{ulem, soul}

\defcitealias{2008AJ....136.2782L}{L08}
\defcitealias{2009ApJ...693..216K}{KMT09a}
\defcitealias{2009ApJ...699..850K}{KMT09b}
\defcitealias{2011ApJ...728...88G}{GK11}
\defcitealias{2013MNRAS.436.2747K}{K13}
\defcitealias{2014ApJ...795...37G}{GD14}
\defcitealias{2014ApJ...790...10S}{S14}
\defcitealias{2016ApJ...822...83B}{BS16}

\newcommand{\kmt}{\citetalias{2009ApJ...699..850K}}
\newcommand{\gk}{\citetalias{2011ApJ...728...88G}}
\newcommand{\krumholz}{\citetalias{2013MNRAS.436.2747K}}
\newcommand{\sternberg}{\citetalias{2014ApJ...790...10S}}
\newcommand{\gd}{\citetalias{2014ApJ...795...37G}}

\begin{document}
\shorttitle{Modeling H$_2$ in low metallicity galaxies}
\shortauthors{Polzin et al.}

\title{Modeling molecular hydrogen in low metallicity galaxies}

\correspondingauthor{Ava Polzin}
\email{apolzin@uchicago.edu}

\author[0000-0002-5283-933X]{Ava Polzin}
\affiliation{Department of Astronomy and Astrophysics, The University of Chicago, Chicago, IL 60637, USA} 

\author[0000-0003-4307-634X]{Andrey V. Kravtsov}
\affiliation{Department of Astronomy and Astrophysics, The University of Chicago, Chicago, IL 60637, USA}
\affiliation{Kavli Institute for Cosmological Physics, The University of Chicago, Chicago, IL 60637, USA}
\affiliation{Enrico Fermi Institute, The University of Chicago, Chicago, IL 60637, USA}

\author[0000-0002-6648-7136]{Vadim A. Semenov}
\affiliation{Center for Astrophysics, Harvard \& Smithsonian, 60 Garden St, Cambridge, MA 02138, USA}

\author[0000-0001-5925-4580]{Nickolay Y. Gnedin}
\affiliation{Department of Astronomy and Astrophysics, The University of Chicago, Chicago, IL 60637, USA}
\affiliation{Kavli Institute for Cosmological Physics, The University of Chicago, Chicago, IL 60637, USA}
\affiliation{Fermi National Accelerator Laboratory; Batavia, IL 60510, USA}

\begin{abstract}
We use a suite of hydrodynamics simulations of the interstellar medium (ISM) within a galactic disk, which include radiative transfer, a non-equilibrium model of molecular hydrogen, and a realistic model for star formation and feedback, to study the structure of the ISM and H$_2$ abundance as a function of local ISM properties. We show that the star formation rate and structure of the ISM are sensitive to the metallicity of the gas with a progressively smoother density distribution with decreasing metallicity. In addition to the well-known trend of the \textsc{HI-H$_2$} transition shifting to higher densities with decreasing metallicity, the maximum achieved molecular fraction in the interstellar medium drops drastically at $Z \lesssim 0.2 \, Z_\odot$ as the formation time of H$_2$ becomes much longer than a typical lifetime of dense regions of the ISM.
We present accurate fitting formulae for both volumetric and projected $f_\mathrm{H_2}$ measured on different scales as a function of gas metallicity, UV radiation field, and gas density. We show that when the formulae are applied to the patches in the simulated galaxy the overall molecular gas mass is reproduced to better than a factor of $\lesssim 1.5$ across the entire range of metallicities and scales. We also show that the presented fit is considerably more accurate than any of the previous $f_{\rm H_2}$ models and fitting formulae in the low-metallicity regime. 
The fit can thus be used for modeling molecular gas in low-resolution simulations and semi-analytic models of galaxy formation in the dwarf and high-redshift regimes. 
\end{abstract}


\section{Introduction} 

The cold, dense tail of the multiphase interstellar medium (ISM) is generally home to cold atomic gas \citep[e.g.,][]{Wolfire.etal.2003} and molecules, such as CO, HCN, H$_2$, etc., which play an important role in the thermodynamics of gas in this phase \citep[e.g.,][]{Omont.etal.2007,Draine.2011.book,Galli.Palla.2013}.  At the same time, molecular gas is one of the very few direct observational probes of this tail \citep[see][for reviews]{Carilli.Walter.2013,2022ARAA..60..319S}. Given that stars also form in high-density regions, empirical studies of molecular gas are intricately tied to studies of how star formation occurs in galaxies \citep[e.g.,][]{2012ARAA..50..531K}. 

Empirically, it is established that a fairly tight relation between surface densities of molecular gas and star formation exists both within individual galaxies and among different galaxies \citep[the molecular Kennicutt-Schmidt relation; e.g.,][]{1989ApJ...344..685K, 1998ApJ...498..541K,Wong.Blitz.2002, Bigiel.etal.2008,2023MNRAS.518.4767B}, the relation that is now well understood theoretically \citep[e.g.,][]{2019ApJ...870...79S}. Observationally, that correlation is tied to CO, but the assumption is that CO traces H$_2$ reasonably well, with the possible exception of extreme starbursts \citep{2001AJ....121..740M, 2012AJ....143..138S, Carilli.Walter.2013,2022EPJWC.26500011M}.

The existence of such a tight correlation was used as a basis for modeling star formation in galaxy formation simulations \citep[e.g.,][]{Robertson.Kravtsov.2008,Gnedin.etal.2009,Jaacks.etal.2013,Christensen.etal.2014} and semi-analytic models \citep[e.g.,][]{Popping.etal.2014} and motivated development of theoretical models of molecular gas  \citep[e.g.,][see \citealt{2018ApJS..238...33D} for a review]{2008AJ....136.2782L, 2009ApJ...693..216K, 2011ApJ...728...88G, 2014ApJ...795...37G,2014ApJ...790...10S}.
However, most existing models of molecular hydrogen gas fraction are calibrated in the relatively high mass, high metallicity regime. The \textsc{HI-H$_2$} transition models that are formulated for lower $Z$ gas are often more complex and have additional  assumptions and tunable parameters \citep{2009ApJ...699..850K, 
2013MNRAS.436.2747K, 2014ApJ...795...37G, 2016ApJ...822...83B}. Furthermore, most models estimate the abundance of H$_2$ assuming chemical equilibrium whereby the process of molecular gas formation is not limited in time.

The low-metallicity regime is different. It is generally expected that the formation time of H$_2$ increases with decreasing metallicity. At the same time, the lifetime of dense regions of the ISM is finite due to a combination of shearing forces and effects of stellar feedback. If the lifetimes of dense ISM regions are shorter than the characteristic H$_2$ formation time, the molecular fraction in low-metallicity gas may never reach high values, which means that stars in such regions must form from the largely atomic gas \citep{2012ApJ...759....9K, Glover.etal.2012b,2016MNRAS.458.3528H}. Indeed, physically 
star formation can occur in purely atomic gas because cooling and other processes driving the formation of star-forming regions are only mildly affected by the presence of molecular gas \citep{Glover.etal.2012}.
This implies that chemical equilibrium models of $f_{\rm H_2}$ that assume no time limit to H$_2$ formation systematically overpredict $f_\mathrm{H_2}$ in low-metallicity gas \citep{2012ApJ...759....9K}. Conversely,  as shown by \citet{Glover.etal.2012b} models that use non-chemical equilibrium calculations of H$_2$ abundance to estimate star formation rate will underpredict star formation rate, if star-forming regions have low molecular fractions but otherwise form stars with a regular efficiency. 

Low metallicities are relevant for modeling the two regimes of galaxy evolution that are at the current frontiers of extragalactic research: the earliest stages of evolution of massive galaxies at $z\gtrsim 5$ and evolution of local dwarf galaxies. It is thus important to examine and calibrate the abundance of molecular gas and star formation efficiency in this low-metallicity regime. Likewise, theoretical models of 
the interstellar medium and star formation in galaxy formation simulations may potentially be tested by contrasting their results with observational estimates of the H$_2$ abundance and star formation in dwarf galaxies \citep[e.g.,][]{2011ApJ...741...12B,2016ApJ...825...12J} and galaxies at high redshifts.

In this paper, we examine the abundance of molecular hydrogen -- the dominant mass component of molecular gas -- using a suite of realistic simulations of a dwarf galaxy's ISM across a wide range of metallicities. These simulations use a generalized star formation prescription that is not based on the local H$_2$ abundance \citep{2021ApJ...918...13S}. Instead, the star formation prescription in the simulations is based on the results of high-resolution magneto-hydrodynamic simulations of star-forming regions \citep{padoan12,semenov16}. Most importantly, the model reproduces the abundance and spatial distribution of molecular and atomic gas in NGC 300, and the observed de-correlation between cold molecular gas and clusters of young stars as a function of scale in this galaxy \citep{2021ApJ...918...13S}. This gives credence to the model as a benchmark that can be used to calibrate star formation and molecular gas abundance in the ISM as a function of its properties. 

We present fitting formulae for both volumetric and projected molecular hydrogen fractions that depend on the gas density, gas metallicity, and local ionizing UV field. We show that the fits reproduce known trends in the location and shape of the HI-H$_2$ transition with metallicity and accurately reproduce both the form of the dependence of molecular fraction on the (volume or column) density and the total molecular gas mass in simulations. In addition, we demonstrate that the structure and behavior of the ISM changes qualitatively when gas metallicity decreases to $\lesssim 0.1Z_\odot$. 

The paper is organized as follows. In Section \ref{sec:sim}, we describe the simulation used to calibrate our $f_\mathrm{H_2}$ models. In Section \ref{sec:fmol_model}, we lay out our simple models for $f_\mathrm{H_2}$ to be used in both a volumetric and projected case, and in Section \ref{sec:results}, we present tests of the accuracy of the models. Finally, in Section \ref{sec:discussion}, we discuss the implications for models of galaxy formation and compare them against existing $f_{\rm H_2}$ models, details of which are presented in the Appendix~\ref{app:model_implement}.

\begin{figure*}[ht!]
\epsscale{1.2}
\plotone{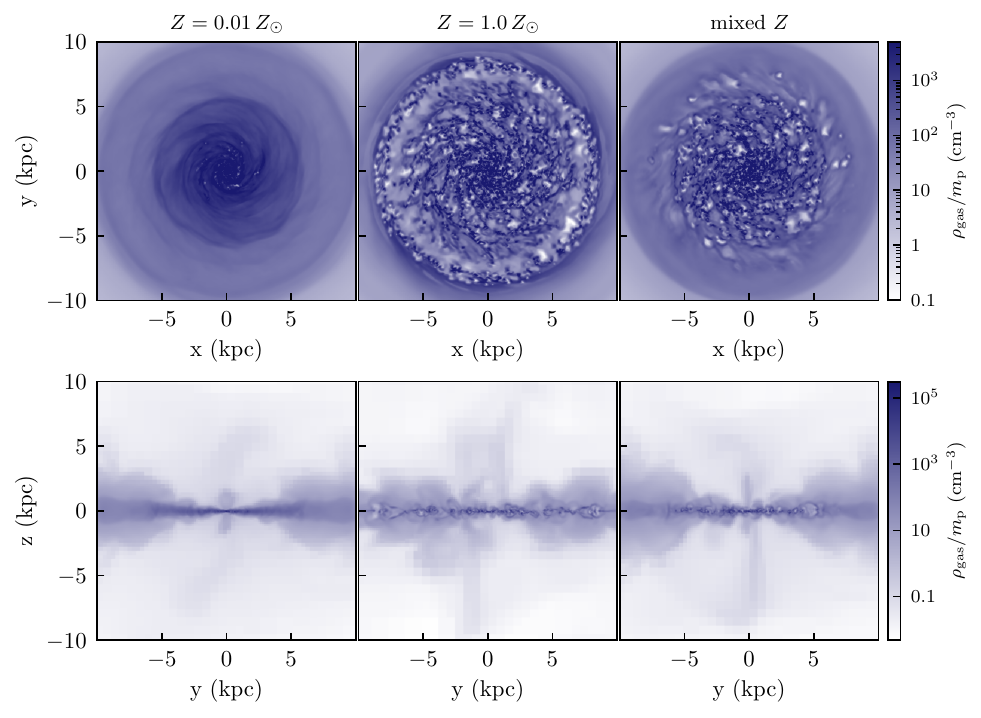}
\caption{Face-on (\emph{top}) and edge-on (\emph{bottom}) gas density slices of our simulated NGC 300-like galaxy in our runs with the lowest and highest $Z$ fixed metallicity as well as our fiducial simulation with variable metallicity. 
\label{fig:disk}}
\end{figure*}

\section{Simulations}
\label{sec:sim}

We conduct our analysis using a suite of simulations of an isolated disk galaxy that is initialized with structural properties similar to those observed in the dwarf galaxy NGC 300 \citep{2021ApJ...918...13S}. We refer the reader to that paper for a detailed description of the simulation setup and implementations of various included physical processes. Below we summarize the key aspects of the simulations. 

The fiducial simulation has been shown to reproduce details of the star formation and atomic and molecular gas distributions in NGC 300, including the observed spatial decorrelation of cold gas with recent star formation as a function of averaging scale \citep[or the ``tuning fork'';][]{2019Natur.569..519K}. 

The simulations are carried out using the ART $N$-body$+$hydrodynamics code \citep{1999PhDT........25K, 2002ApJ...571..563K, 2008ApJ...672...19R, 2011ApJ...728...88G} with self-consistent modeling of radiative transfer \citep[RT;][]{2014ApJ...793...29G} and non-equilibrium abundance of molecular hydrogen coupled to the local UV radiation field \citep[using the 6-species model described in the Appendix of][]{2011ApJ...728...88G}. With the inclusion of RT and a realistic ISM structure shaped by star formation and feedback, as well as the simulation's maximum resolution of $\sim 10$ pc (the average grid cell size is $\sim 22$ pc when $n_\mathrm{H} \ge 0.1\,\rm cm^{-3}$), the simulation offers a highly realistic model for the formation/destruction of molecular hydrogen gas. 

\begin{deluxetable}{lcccc}
\tablecaption{Basic properties of the snapshots (within 15 kpc of the simulation center), which were fit to construct our models. Each version of the simulation was run for sufficient time that $M_\mathrm{H_2}$ was not evolving significantly between snapshots. \label{tab:props}}
\tablewidth{0.5\linewidth}
\tablehead{Run & Time/Myr & $M_\mathrm{H_2}$/$M_\odot$ & $M_\mathrm{HI}$/$M_\odot$ & $M_\mathrm{gas}$/$M_\odot$}
\startdata
$Z = 0.01 \, Z_\odot$ & 900 & $1.4 \times 10^4$& $1.2 \times 10^9$ & $2.0 \times 10^{9}$\\
$Z = 0.03 \, Z_\odot$ & 921 & $4.8 \times 10^4$& $1.2 \times 10^{9}$ & $2.0 \times 10^{9}$\\
$Z = 0.1 \, Z_\odot$ & 880 & $6.4 \times 10^5$ & $1.1 \times 10^{9}$ & $2.0 \times 10^{9}$\\
$Z = 0.2 \, Z_\odot$ & 890 & $3.5 \times 10^6$ & $1.1 \times 10^{9}$ & $2.0 \times 10^{9}$\\
$Z = 0.3 \, Z_\odot$ & 841 & $7.3 \times 10^6$ & $1.1 \times 10^9$ & $2.0 \times 10^{9}$\\
$Z = 0.6 \, Z_\odot$ & 801 & $2.2\times 10^7$ & $1.1\times 10^9$ & $1.9 \times 10^{9}$\\
$Z = 1.0 \, Z_\odot$ & 821 & $4.3 \times 10^7$ & $1.1\times 10^9$ & $2.0 \times 10^{9}$\\
mixed $Z$\tablenotemark{a} & 901 & $1.9 \times 10^7$ & $1.1 \times 10^9$ & $1.9 \times 10^{9}$\\
\enddata
\tablenotetext{a}{For our fiducial mixed metallicity run, the grid cells range in $Z$ from $10^{-20}$ to $\sim4\, Z_\odot$, with a median metallicity of $\sim 0.4\, Z_\odot$ in higher density ($n_\mathrm{H} \ge 0.1$ cm$^{-3}$) gas.}
\end{deluxetable}

Star formation in the simulation is not tied to $f_\mathrm{H_2}$ but follows the prescription introduced in \citet{semenov16}. This implementation uses a dynamical model for unresolved turbulence to predict locally variable star formation efficiency instead of assuming a constant tunable value. As was shown in \citet{2017ApJ...845..133S, 2019ApJ...870...79S, 2021ApJ...918...13S}, modeling star formation efficiency based on local properties of turbulence is important for reproducing the linear molecular KS relation on kiloparsec scales and the spatial decorrelation between young stars (UV sources) and molecular gas regions on sub-kiloparsec scales. It was also shown that this model can reproduce star formation and molecular gas properties both in Milky Way-sized galaxies and in a dwarf galaxy like NGC 300. 
Results of the analyses presented in this paper should therefore be generally applicable to a wide range of regular galaxies with similar chemical and physical properties. We note, however, that the ISM in strongly starbursting galaxies can have a considerably different density distribution and both the abundance of molecular gas and star formation may behave differently in these environments compared to predictions of our model. 

In our fiducial simulation, the metallicity of gas is initialized to have the radial profile similar to the metallicity profile observed in NGC300, after which metallicity is evolved self-consistently. The final snapshot that we use in our analysis contains cells with gas metallicity ranging from effectively $0$ to $4.2 \, Z_\odot$ (the former corresponds to the halo gas, while the latter gas is in the regions newly enriched by supernova ejecta).

In order to examine the role of metallicity in determining the molecular gas fraction and star-forming gas fraction, we ran a suite of seven re-simulations of the same galaxy, in which gas metallicity is fixed to values approximately evenly distributed in $\log_{10}Z$: $Z = 0.01,\, 0.03,\, 0.1,\, 0.2,\, 0.3,\, 0.6,\, 1.0\, Z_\odot$. Note that, although gas metallicity is fixed in these runs, all other processes are modelled in the same way. In particular, radiative transfer is performed and UV field varies spatially, reflecting the distribution of sources and absorbing gas. We use the variation of the UV flux within a run to study the dependence of the molecular fraction on this flux at a given gas metallicity. 
Details of these runs are presented in Table \ref{tab:props}.

Figure \ref{fig:disk} presents face-on and edge-on views of the gas distribution in three of these runs. The gas density distribution varies substantially with metallicity, with a more homogeneous density distribution at low $Z$ and a more flocculent density distribution at higher $Z$. This trend is also apparent in the fiducial run with a non-uniform metallicity distribution shown in the right column, in which inner high-$Z$ regions are similar to the $Z=1\, Z_\odot$ run, while the outer lower-$Z$ regions have a much smoother gas distribution. 

\begin{figure}
\epsscale{1.2}
\plotone{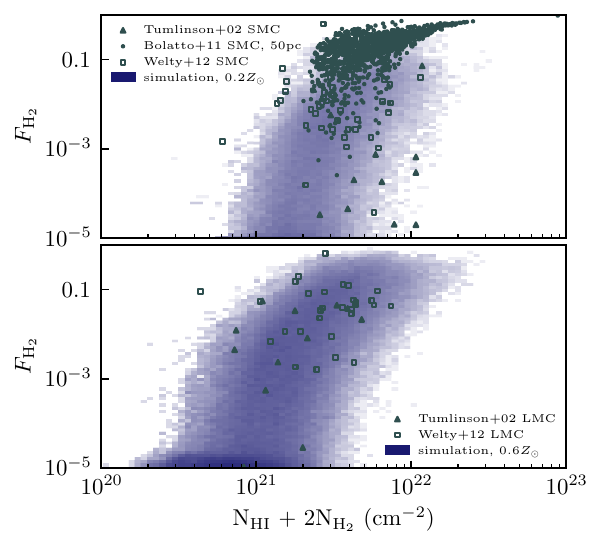}
\caption{Molecular hydrogen gas fraction in projected patches with $S = 10$ pc as a function of gas column density. We compare the $Z = 0.2 \, Z_\odot$ and $Z = 0.6 \, Z_\odot$ simulation runs to observations of $F_\mathrm{H_2}$ in the Small Magellanic Cloud (SMC; \emph{top}) and Large Magellanic Cloud (LMC; \emph{bottom}) respectively. 
\label{fig:obs_compare}}
\end{figure}

We compare the projected molecular hydrogen fraction, $F_\mathrm{H_2} \equiv N_{\rm H_2}/(N_{\rm HI}+2N_{\rm H_2})$, from the single $Z$ runs of the simulation to observations of $F_\mathrm{H_2}$ in the Large and Small Magellanic Clouds \citep{2002ApJ...566..857T, 2011ApJ...741...12B, 2012ApJ...745..173W} in Figure \ref{fig:obs_compare}. We select the single $Z$ run closest to the gas metallicities of the LMC and SMC, adopting our $0.6 \, Z_\odot$ run as an analog for the LMC and our $0.2 \, Z_\odot$ run as an analog for the SMC. The location of the HI-H$_2$ transition is in good agreement between the observations and the simulation. 

Although we do not exactly match the high $F_\mathrm{H_2}$ observations of the SMC from \citet{2011ApJ...741...12B} in our $0.2\, Z_\odot$ run, there are two effects that likely contribute to this discrepancy. The novel method for measuring $\Sigma_\mathrm{H_2}$ used in \citet{2011ApJ...741...12B} does not fully distinguish between H$_2$ and cold HI, potentially yielding a slightly higher molecular fraction that is ultimately more reflective of the fraction of cold neutral gas. Additionally, even though the highest resolution of our simulation grid cells is 10 pc, the effective resolution is several times this, which means that we are not sensitive to features below this effective resolution scale. It is then possible that the less prevalent high density, highest $F_\mathrm{H_2}$ regions are averaged to somewhat lower molecular fraction.

\section{Modeling the molecular gas fraction}
\label{sec:fmol_model}

In this section, we present two versions of fitting formulae suitable for application in different regimes: fits to a volumetric $f_{\rm H_2}$ fitted to the cell-by-cell distribution of the molecular gas fraction in simulations, which can be used in high-resolution simulations (Section~\ref{sec:fmol_vol}), and fits to projected molecular fraction $F_{\rm H_2}$, fitted to projected 2D maps averaged on different spatial scales, which can be used in low-resolution simulations and semi-analytic models (Section~\ref{sec:fmol_proj}).

\subsection{Volumetric $f_{\rm H_2}$ model}
\label{sec:fmol_vol}

We model the molecular hydrogen fraction as a function of hydrogen gas density, metallicity, and UV field strength in individual simulation grid cells. 
The functional form of $f_{\rm H_2}$ fit is motivated by the fact that we expect $f_{\rm H_2}$ to exhibit a fairly sharp transition at a certain density or column density and saturate at values close to some maximum value $f_{\rm H_2,max}$. We thus choose a sigmoid-like function:
\begin{equation}
\label{eq:fh2_vol}
    f_\mathrm{H_2} = \frac{f_\mathrm{H_2, max}}{1 + \exp(-x + \ln f_\mathrm{H_2, max} + 7.42)}.
\end{equation}
This form allows us to account for the fact that the maximum possible molecular hydrogen fraction $f_{\rm H_2,max}$ varies as a function of metallicity. We parameterize this dependence as

\begin{equation}
    \label{eq:fmax}
    f_\mathrm{H_2,max} = [1 + 2\,(1 - f_{\rm m})/f_{\rm m}]^{-1}
\end{equation}
where
\begin{equation}
    f_{\rm m} = 1 - \exp(-Q)
\end{equation}
and
\begin{equation}
    Q = 6 R_0\,\left(\frac{Z}{0.2}\right)^{1.3} n_\mathrm{H}
\;\mathrm{Myr}.
\end{equation}
Here $R_0 = 3.5 \times 10^{-17}$ cm$^3$ s$^{-1}$ is the rate of H$_2$ formation on dust grains \citep[see][]{2008ApJ...680..384W}.

We define $x$ as 
\begin{equation}
    x \equiv g(Z) \ln \frac{n_\mathrm{H}}{n_\mathrm{tr}}
    \label{eq:fh2trans}
\end{equation}
where
\begin{equation}
    g(Z) = 7.6\, Z^{0.25}.
\end{equation}
The functional form and parameter values in these equations were chosen so that the average trend of $f_{\rm H_2}$ with gas density in the simulation and the maximum values of $f_{\rm H_2}$ are reproduced. Equation~\ref{eq:fh2trans}
accounts for the dependence of the location and shape of the \textsc{HI-H$_2$} transition on UV field strength and metallicity. 
The density at which this transition occurs is set by the value of $n_\mathrm{tr}$, while the metallicity-dependent pre-factors are responsible for the changing slope of $f_\mathrm{H_2}$ vs. $n_\mathrm{H}$. The value of $n_\mathrm{tr}$ can be measured from the simulation and fit directly.

\begin{figure*}[ht!]
\epsscale{1.2}
\plotone{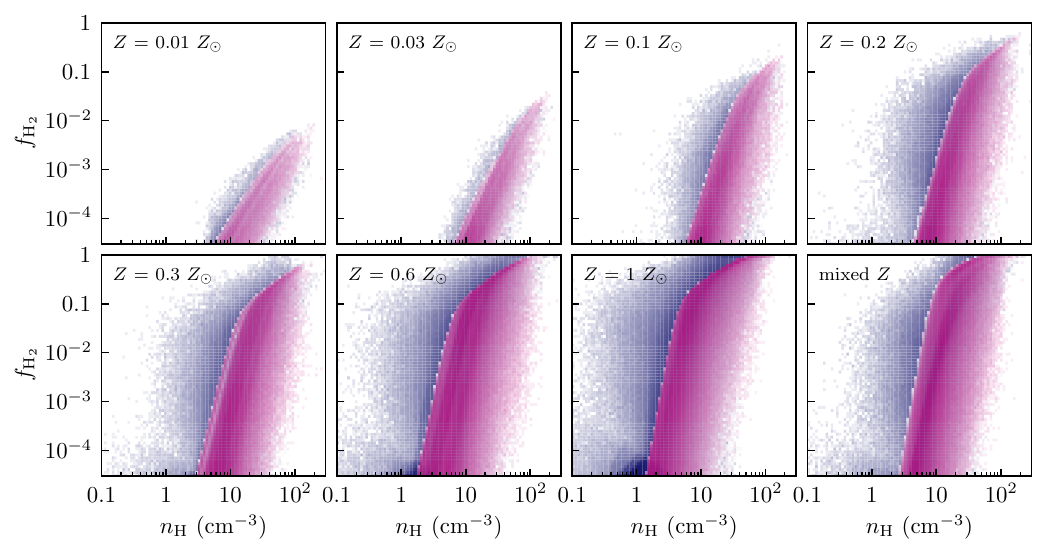}
\caption{Molecular hydrogen gas fraction ($f_{\mathrm H_2}$) in computational grid cells of different hydrogen gas density, $n_\mathrm{H}$, for each run in our suite. The values in simulation cells are shown in blue, with the model described in Equation \ref{eq:fh2_vol} overplotted in pink. 
The model explicitly captures the behavior of $f_\mathrm{H_2}$ at high $n_\mathrm{H}$ owing to the metallicity- and density-dependent cap value $f_\mathrm{H_2, max}$ (see Equation~\ref{eq:fmax}).
The model was not designed to reproduce the tail of high $f_{\rm H_2}$ at lower densities for the reasons discussed in Section~\ref{app_compare}.
\label{fig:fh2_vol}}
\end{figure*}

Given the very low molecular gas fraction at low metallicities, the parameterization using $n_{1/2}$ -- the density at which $f_{\rm H_2}=0.5$  in \citet{2011ApJ...728...88G} and \citet{2014ApJ...795...37G} at higher metallicities does not work in our lowest-$Z$ runs as $f_\mathrm{H_2}$ never reaches 0.5 (see Figure \ref{fig:fh2_vol}). Instead, we define $n_\mathrm{tr}$ as the hydrogen number density (cm$^{-3}$) at which the molecular hydrogen fraction is $5\times10^{-4}$, which characterizes the transition even in this low metallicity regime.

\begin{figure}
\epsscale{1.2}
\plotone{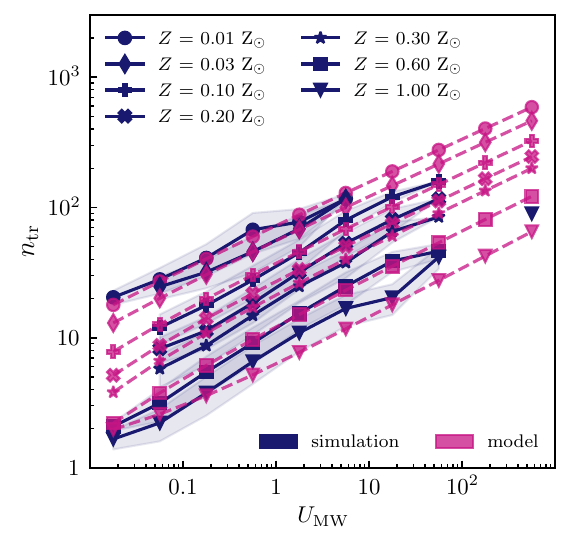}
\caption{Values of $n_\mathrm{tr}$ estimated for each metallicity  and $U_{\rm MW}$ bin as a function of $U_\mathrm{MW}$ are shown for each of the seven runs with fixed metallicities (blue symbols connected by solid lines). The shaded region corresponds to the 16th and 84th percentile of $n_\mathrm{tr}$ in each bin. The $n_\mathrm{tr}$ model results as a function of metallicity and $U_\mathrm{MW}$ (Equation \ref{eqn:cellbycell}) are shown by magenta symbols connected by the dashed lines. 
\label{fig:fit_compare}}
\end{figure}

Figure \ref{fig:fh2_vol} shows that the atomic-to-molecular transition occurs at lower $n_\mathrm{H}$ for higher metallicities, while Figure \ref{fig:fit_compare} shows that $n_\mathrm{tr}$ behaves like a power law with respect to both $U_\mathrm{MW}$ and $Z$. This power law at each discrete metallicity for binned values of $U_\mathrm{MW}$ on a volumetric cell-by-cell basis can be approximated by
\begin{equation}
    n_\mathrm{tr}(D, U_\mathrm{MW}) = b\,(U_\mathrm{MW}) -a\,(D, U_\mathrm{MW}) \log_{10}D +  c\,(D) \label{eqn:cellbycell}
\end{equation}
Here $U_\mathrm{MW}$ is the free-space\footnote{Free-space UV flux is the flux at a given location not attenuated by {\it local} extinction.
 \citet{2009ApJ...693..216K},
for example, use free-space flux to mean the flux incident on molecular clouds. In our simulations, the radiative transfer calculations do not include absorption by H$_2$ lines
and thus do not model radiation field self-consistently inside molecular-rich regions and have to rely on the subgrid model. In this case the free space flux is the flux returned by the radiative transfer solver and has the physical meaning of the incident field on the molecular gas.}
UV flux relative to the MW value $U_{\rm MW}=J_{1000}/J_{\rm MW}$, where $J_{1000}$ is the interstellar UV flux at 1000\AA\, $J_{\rm MW}=10^6\,\rm photons\,cm^{-2} \,s^{-1} \,ster^{-1} \,eV^{-1}$ \citep[][]{Draine.1978,Mathis.etal.1983}, and $D$ is the dust-to-gas ratio, which we assume to be equal to the unnormalized mass fraction of heavy elements in the gas. 

We then determine fit parameters $a$ and $b$ as a function of $U_\mathrm{MW}$ using a simple least squares fit of simulation results:
\begin{equation}
    a = 34.7\, U_\mathrm{MW}^{0.32} - 2.25\,\left(\frac{D}{0.0199}\right)^{0.3} \notag
\end{equation}
\begin{equation}
    b = -53.9 \,U_\mathrm{MW}^{0.31} \notag
\end{equation}
The strength of the $D$-dependence of $n_\mathrm{tr}$ becomes weaker at higher metallicities for all $U_\mathrm{MW}$. To reflect this saturation of the metallicity dependence at near-solar metallicities, we add a correction term assuming that solar metallicity corresponds to mass fraction of 0.0199:
\begin{equation}
c(D) = \frac{D}{0.2 \times 0.0199}. 
\label{eq:cD}
\end{equation}

To avoid non-physical, negative $n_\mathrm{tr}$ at high $Z$ and very low $U_\mathrm{MW}$, the floor of $n_\mathrm{tr}$ can be set explicitly. We choose a minimum value of $n_\mathrm{tr} = 0.1$ cm$^{-3}$ given that we anticipate very little cold, dense molecular hydrogen gas at $n_\mathrm{H} \lesssim 0.1$ cm$^{-3}$, but this can be set even lower without affecting the accuracy of the overall model. The results of this fit to $n_\mathrm{tr}$ are shown in Figure \ref{fig:fit_compare} overplotted on the measured location of this transition.

Equations~\ref{eq:fh2_vol}-\ref{eq:cD} can be used to estimate $f_{\rm H_2}$ in the high-density gas (see Figure \ref{fig:fh2_vol}), as this gas constitutes most of gas mass in galaxies. The molecular fraction in the low-density unshielded gas does not require a fitting formula and can be obtained by simply equating the H$_2$ formation and photo-dissociation rates \citep[see, e.g., Section 3.2 in][]{2008ApJ...680..384W}: 
\begin{equation}
   f_{\rm H_2}=\frac{2n_{\rm H}R_0}{U_{\rm MW}I}, 
\end{equation}
where $R_0\approx 3.5\times 10^{-17}(D/0.019)\,\rm cm^3\,s^{-1}$ is the rate of H$_2$ formation on dust grains assumed here to scale linearly with $D$, $n_{\rm H}$ is the number density of hydrogen nuclei, and $I=4.7\times 10^{-11}\,\rm s^{-1}$ is the unshielded photodissociation rate in the local interstellar UV field.

\subsection{Model for projected molecular fraction}
\label{sec:fmol_proj}

In observations, low-resolution simulations, and semi-analytic models one often needs to work with the projected mass densities and we thus present fitting formulae for projected molecular fraction below. 
To distinguish it from the volumetric one we will denote the projected fraction as
\begin{equation}
F_{\rm H_2} = \frac{\Sigma_{\rm H_2}}{\Sigma_{\rm H_2}+\Sigma_{\rm HI}}.
\end{equation}

To obtain projected molecular fractions $F_{\rm H_2}$ on different spatial scales in the simulations, we use the face-on projection of the simulated galaxy with gas properties binned on grids with physical cell sizes 10 pc to 1 kpc. The gas surface (column) density is computed simply as the gas mass (atom number) in each bin divided by its area. 
The UV flux and metallicity in each bin are estimated as the gas-density weighted averages of $U_\mathrm{MW}$ and $Z$ in the computational cells enclosed in a given bin. 

\begin{figure*}
\epsscale{1.2}
\plotone{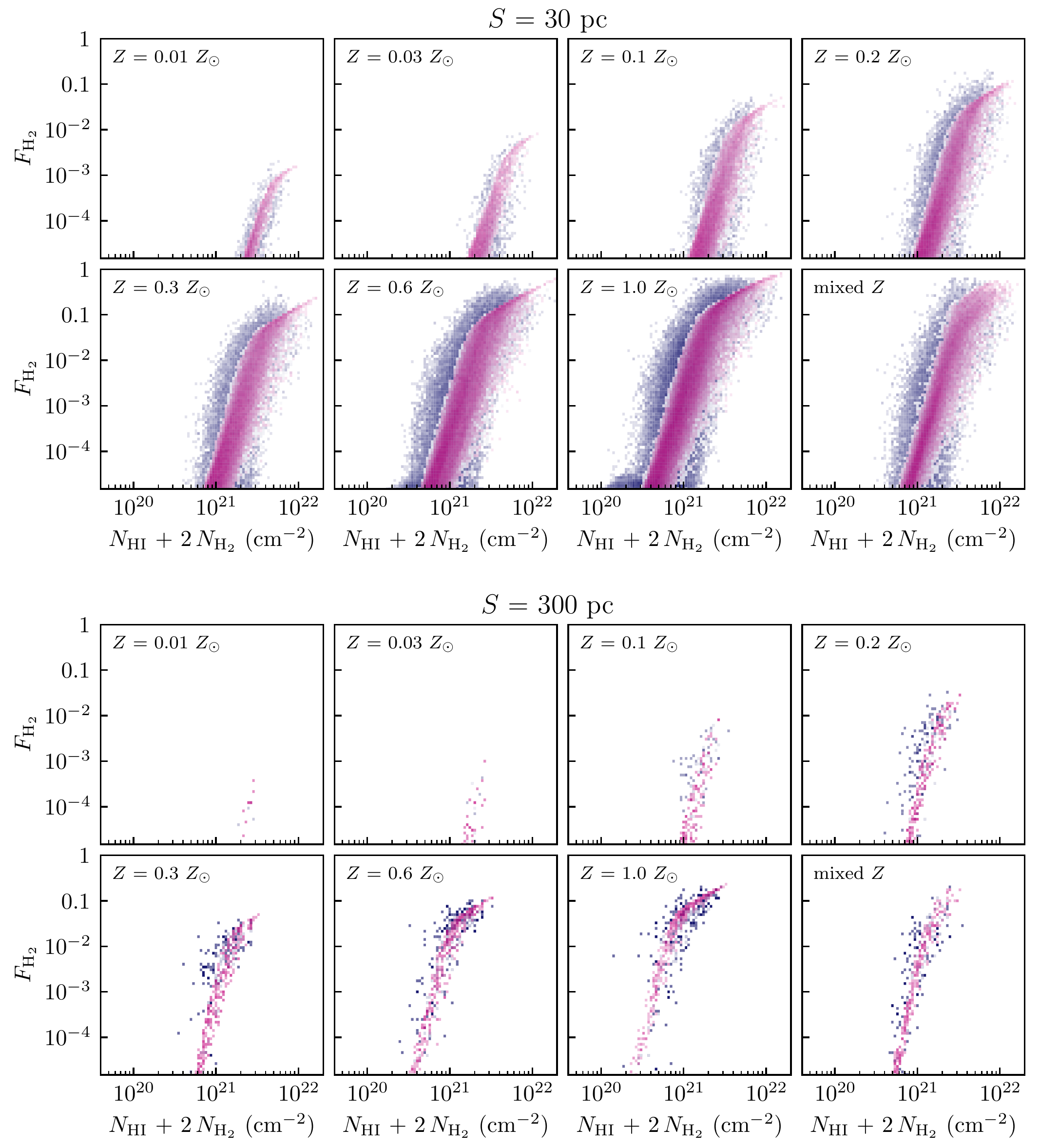}
\caption{Molecular hydrogen gas fraction ($F_{\mathrm H_2}$) in projected patches of size $S$ of different hydrogen gas column density, $N_\mathrm{HI} + 2\,N_\mathrm{H_2}$, for each run in our simulation suite at two representative scales, $S=30$ pc and $S=300$ pc. The simulation values are shown in blue, while the model described in Equation \ref{eq:fh2_vol} is overplotted in pink. 
\label{fig:fh2_proj}}
\end{figure*}

Projected molecular fractions as a function of column density, $F_\mathrm{H_2}$ ($N_\mathrm{H}$), are shown in Figure \ref{fig:fh2_proj} for two representative averaging scales -- 30 pc and 300 pc. Similarly to the volumetric molecular fractions, we use the sigmoid-like functional form of the model for $F_{\rm H_2}$:
\begin{equation}
        F_\mathrm{H_2} \approx \frac{F_\mathrm{H_2, max}}{1 + \exp(-x + \ln F_\mathrm{H_2, max} +8.71)},  
\label{eq:fh2_proj}
\end{equation}
where
\begin{equation}
    F_\mathrm{H_2, max} = [1 + 2(1 - F_{\rm m})/F_{\rm m}]^{-1},
\end{equation}
and
\begin{equation}
    F_{\rm m} = 1 - \exp(-Q),
\end{equation}
\begin{equation}
    Q = 3 R_0 \, \left(\frac{Z}{0.1}\right)^{1.3} \, \frac{N_{\mathrm H}}{4.63 \times 10^{20} \, \mathrm{cm}}\  \mathrm{Myr}, 
\end{equation}
and where, as before, $R_0 = 3.5 \times 10^{-17}$ cm$^3$ s$^{-1}$ \citep[see][]{2008ApJ...680..384W}, and $x$ in Equation~\ref{eq:fh2_proj} is
\begin{equation}
    x = g(Z,S)\,\ln\frac{N_{\mathrm H}}{N_{\mathrm tr}},
\end{equation}
where
\begin{eqnarray}
    g(Z,S) = 1 &+& 1.35\left(\frac{Z}{0.01}\right)^{-0.25}\left(\frac{S}{10\, \mathrm{pc}}\right)^{0.6}\nonumber\\
    &+& 3.4\,\left(\frac{Z}{0.6}\right)^{0.02},
\end{eqnarray}
and $S$ is the resolution of the projected map from the simulation, i.e. the scale on which surface densities and fractions are averaged.  

The transition column density, $N_\mathrm{tr} (Z, U_\mathrm{MW}, S)$, is defined as the median column density of all cells with molecular hydrogen fractions between $5\times 10^{-5}$ and $5\times 10^{-4}$. This region in the $F_\mathrm{H_2}-N_\mathrm{H}$ parameter space was chosen because $F_\mathrm{H_2} (N_\mathrm{H})$ is increasing sharply with increasing column density and $F_\mathrm{H_2}$ range is sufficiently low to be used at very small metallicities. As in the volumetric model, the location of the transition is set by $N_\mathrm{tr}$ (Figure \ref{fig:nmed}), while the slope of $F_\mathrm{H_2} (N_\mathrm{H})$ is set by the pre-factors, which have a weak dependence on the scale $S$ (in parsecs) and metallicity, $\zeta=\log_{10}Z$:

\begin{figure*}
\epsscale{1.2}
\plotone{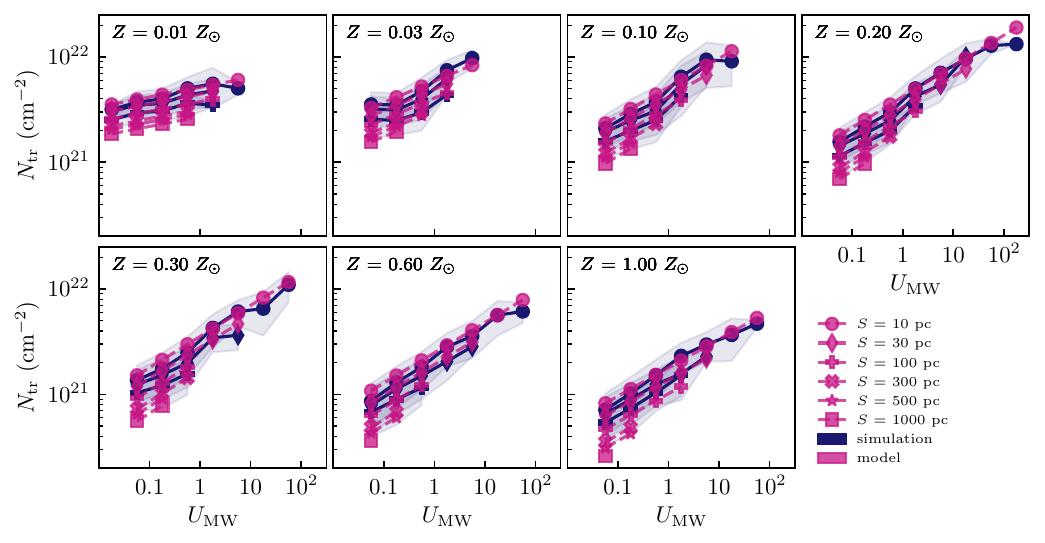}
\caption{The median column density for which the molecular fraction is between $5\times10^{-5}$ and $5\times10^{-4}$, which marks the location of the \textsc{HI-H$_2$} transition, as a function of $U_\mathrm{MW}$ and $S$ for each of our single metallicity runs. The measured values are shown by the dark blue symbols connected by lines. Given that we only fit explicitly for averaging scales $S$ between 10 and 100 pc due to the number of available grid cells, we do not include the measured values for scales between 300 and 1000 pc. 
The model results as a function of $U_{MW}$, $Z$, and $S$ (Equation \ref{eq:proj}) are shown by the magenta symbols connected by dashed lines.
\label{fig:nmed}}
\end{figure*}

\begin{equation}
    N_\mathrm{tr} = U_\mathrm{MW}^{w(\zeta)}\, 10^{y(\zeta, S)}\, N_\mathrm{corr}(\zeta, S)\label{eq:proj}
\end{equation}
\begin{equation}
    w = 0.27 - 0.01\left[9.25 \,\zeta^2 + 9.64 \, \zeta\right], \notag
\end{equation}
\begin{equation}
    y = y_\mathrm{norm}\, \exp\left(-0.5 \left[\frac{\zeta + 1.5}{6.84}\right]^2\right) \notag
\end{equation}
\begin{equation}
\begin{array}{l}
     y_\mathrm{norm} = 21.96 - 0.19\,\log_{10} S.
     \notag 
\end{array}
\end{equation}
As for the volumentric model, the functional forms and their parameter values are chosen so that equation~\ref{eq:fh2_proj} reproduces the mean trend of $F_{\rm H_2}$ as a function of $N_{\rm H}$ in the simulations for each metallicity, UV flux and scale. 

$N_\mathrm{corr}$ is the correction factor introduced to address two distinct limitations of our original fit to the measured median column density of the \textsc{HI}-\textsc{H$_2$} transition. Due to a paucity of grid cells at larger scales ($S > 100$ pc), we only explicitly fit for $N_\mathrm{tr}$ in the $S = 10$, 30, and 100 pc cases, which means that our parameterization does not account for the behavior on larger spatial scales. In addition, given that $F_\mathrm{H_2}$ is defined assuming that the transition occurs at $1.65\times 10^{-4}$ as defined by $N_\mathrm{tr}$, while the median molecular fraction between $5\times10^{-5}$ and $5\times10^{-4}$ varies subtly with scale and metallicity. We capture these effects using the following functional form:
\begin{equation}
N_\mathrm{corr} = 1 - 0.13\,\log_{10}\left(\frac{Z}{0.1}\right)\, \log_{10}\left(\frac{S}{10\,\rm pc}\right).
\label{eq:ncorr}
\end{equation}

\section{Comparisons of fits to simulation}
\label{sec:results}

\subsection{The volumetric $f_{\rm H_2}$ fit}
\label{app_compare}

The validity and accuracy of the volumetric $f_{\rm H_2}$  fitting formulae (Equations \ref{eq:fh2_vol}-\ref{eq:cD}) can be guaged by comparing the $f_\mathrm{H_2}$ according to the fit to the simulation $f_\mathrm{H_2}$ for all grid cells in the simulation with molecular fractions larger than $10^{-5}$. We do this for all fixed metallicity runs and for the fiducial run with non-uniform metallicity, which was not used in deriving the fit. Note also that the latter includes cells with metallicities outside of the range within which we calibrated the model. 

Figure \ref{fig:fh2_vol} shows a good qualitative agreement between the molecular fractions produced by the model fits and the simulation results for each run. Because we impose a strict $f_\mathrm{H_2, max}$ condition, the model $f_{\rm H_2}$ distribution has a sharp boundary at the highest fraction values at each density. 
In principle, one can introduce scatter around the model relations to reproduce the tail of high $f_{\rm H_2}$ cells at small densities. However, we do not think it is worthwhile for two reasons. First, as we show next, the presented fit accurately recovers molecular mass, $M_\mathrm{H_2}$, for the galaxy at all metallicities. This means that the molecular mass in this tail is fairly small. Second, the bulk of the high $f_{\rm H_2}$ gas at low densities could be due to numerical diffusion of radiation and molecular gas around star-forming regions and thus may be a non-equilibrium artifact. 

The accuracy of the fit in reproducing the molecular content of the ISM in the simulated galaxy can be assessed by comparing the total H$_2$ mass estimated with 
when the molecular fraction in each computational cell is estimated using the fit to the actual mass in the simulation. The left panel of Figure~\ref{fig:acc} shows the ratio $M_\mathrm{H_2,sim}/M_\mathrm{H_2,mod}$ as a function of metallicity for the volumetric model as magenta open circles and indicates that the fit is accurate in predicting H$_2$ mass to  $\lesssim 25\%$. 

To test whether results depend on the time snapshot of the simulation we estimated the ratio of the actual-to-model-predicted molecular mass at a series of different 
simulation snapshots. These estimates are shown in Figure~\ref{fig:time_compare}, which shows that generally, the fit accuracy is similar at most snapshots, except for a single snapshot where the ratio increased to $\approx 1.8$ one where likely non-equilibrium processes related to a local starburst changed the ISM significantly. 

The figure shows that variation of the mass ratio from snapshot to snapshot increases significantly in lower metallicity runs. This is likely related to the rapidly increasing H$_2$ formation time with decreasing metallicity. 
Indeed, using $t_\mathrm{chem} = 105 \, (Z/Z_\odot)^{-1}\, (n_\mathrm{H}/\mathrm{cm^{-3}})^{-1}\,(10/f_c)$ Myr from \citet{2012ApJ...759....9K} and assuming a clumping factor $f_c = 10$ due to turbulence on the scale of molecular clouds \citep[see Appendix A.7 of][]{2011ApJ...728...88G}, the H$_2$ formation time for gas of $Z = 0.01\, Z_\odot$ and number density $n_\mathrm{H} = 50$ cm$^{-3}$ should be 210 Myr, while it is only $\approx 2.1$ Myr for $Z = Z_\odot$ gas of the same density. Given that the typical lifetime of a molecular cloud is significantly shorter, $\sim5-15$ Myr \citep{2017ApJ...845..133S}, it is 
not surprising that $f_\mathrm{H_2}$ is generally suppressed in low-$Z$ gas ($Z \lesssim 0.1\,Z_\odot$), where the time for the gas to reach chemical (HI-H$_2$) equilibrium is longer than the timescales on which molecular clouds persist without disruption. 
In low-metallicity runs, H$_2$ abundance is much more susceptible to disruption of individual star-forming regions (which are also fewer), which leads to larger variations of the H$_2$ abundance. 

We note that the fit should not to be extrapolated to the \textit{zero} metallicity case. In practice, however, a very small value of metallicity should return reasonable results.

\subsection{The projected $F_{\rm H_2}$ fit}
\label{sec:proj_test}

Figure \ref{fig:fh2_proj} shows good agreement between the projected molecular fraction, $F_{\rm H_2}$, estimated using fit (equations~\ref{eq:fh2_proj}-\ref{eq:ncorr}) and simulation results across metallicities \textit{and} across averaging scales. We show this explicitly for two representative scales of 300 pc and 30 pc, both of which are fairly well-populated by simulation grid cells with $f_\mathrm{H_2} \ge 10^{-5}$ even at the lowest metallicities. As with the volumetric fit, the imposed maximum fraction $F_\mathrm{H_2, max}$ results in a hard upper boundary on $F_\mathrm{H_2}(N_\mathrm{H})$ (see Section \ref{app_compare}). Also, the scatter of $F_{\rm H_2}$ in the model at a given number density is somewhat smaller than in the simulations. As we argued in the discussion of the volumetric model comparisons above, the additional scatter can be added to the model, but the amount of molecular gas associated with the tails of the distribution is fairly small. 

\begin{figure*}
\epsscale{1.1}
\plotone{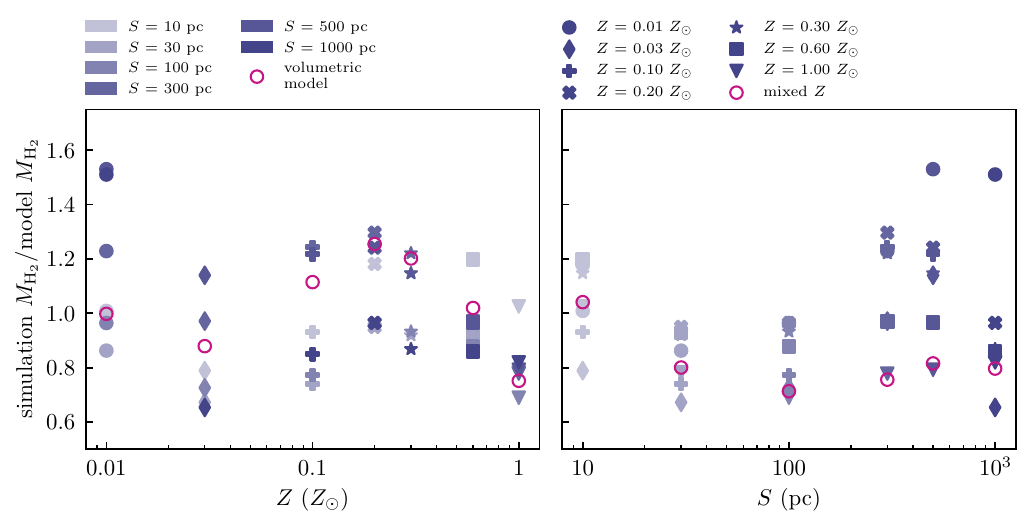}
\caption{The ratio of H$_2$ mass measured in the simulations using the  grid cells with $f_\mathrm{H_2} \ge 10^{-5}$ to the H$_2$ mass predicted by the model using the densities, metallicities, and UV fluxes in the individual simulation cells as a function of metallicity for both the volumetric case (magenta, \emph{left}) and the scale-dependent projected measurements. Notably, the model is  accurate to better than a factor of $\approx 1.5$ in most cases for both the fixed metallicity runs and the fiducial run with non-uniform metallicity (magenta, \emph{right}) across a range of scales. 
\label{fig:acc}}
\end{figure*}

\begin{figure}
\epsscale{1.2}
\plotone{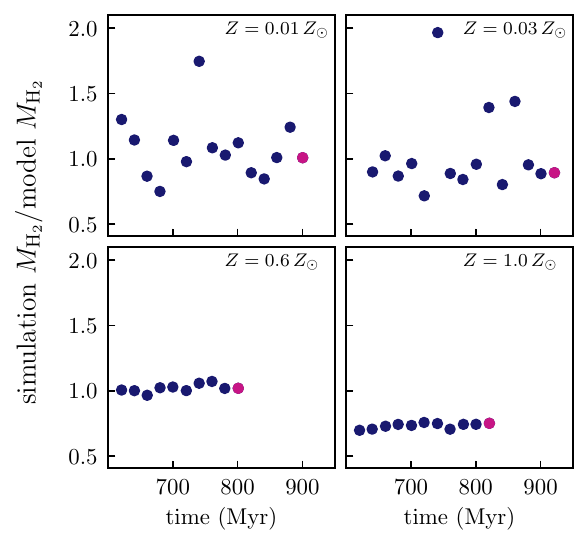}
\caption{The ratio of the H$_2$ mass in the simulation and the H$_2$ mass inferred from the volumetric model (Equation \ref{eq:fh2_vol}) as a function of snapshot time. The most advanced snapshot used in calibrating our model is shown by the magenta point. We show the evolution of this ratio for the two lowest metallicity and two highest metallicity runs we use in this work. The figure indicates that results of the test shown in Figure~\ref{fig:acc} are not sensitive to the specific output used.
\label{fig:time_compare}}
\end{figure}

Indeed, Figure~\ref{fig:acc} shows that the total molecular gas mass estimated using the fit formulae is accurate to better than a factor of $\lesssim 1.5$across the full probed range of metallicities and averaging scales.

\section{Discussion}
\label{sec:discussion}
\subsection{Implications for galaxy formation modeling}

In simulations and analytical models of galaxies, molecular gas is sometimes used as a proxy for star-forming gas, which is motivated by the observed constant depletion times of molecular gas in normal (non-starburst) galaxies of metallicities $\approx 0.1-1Z_\odot$ (see Introduction). 

\begin{figure}[ht!]
\epsscale{1.2}
\plotone{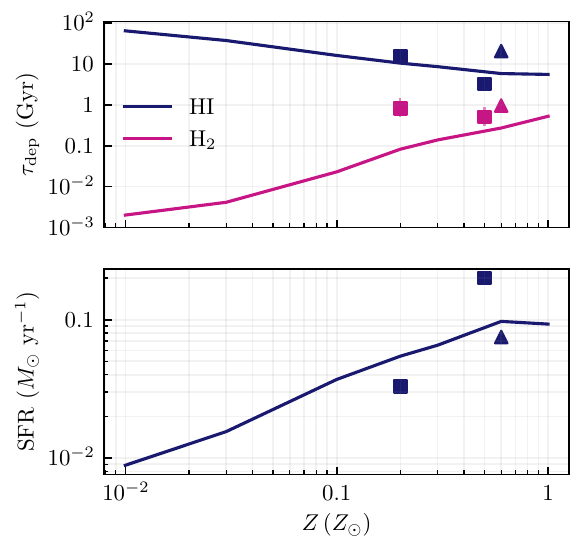}
\caption{The depletion time ($\tau_\mathrm{dep} = M_\mathrm{gas}/\mathrm{SFR}$) as a function of metallicity and gas species relative to the star formation rate (SFR) as a function of metallicity. We use $M_\mathrm{gas}$ within 5 kpc of the simulation center. While $\tau_\mathrm{dep,H_2}$ varies by a factor of $260$ between $Z = 0.01 \, Z_\odot$ and $1 \, Z_\odot$, the star formation rate (defined here by the mass of stars formed over the last 10 Myr) only varies by a factor of $11$. This is indicative of the fact that the star formation in low metallicity galaxies is not directly tied to H$_2$ abundance. We also include the inferred SFR and H$_2$ and \textsc{HI} $\tau_\mathrm{dep}$ in the Large and Small Magellanic Clouds from \citet{2016ApJ...825...12J} as squares, which we correct by a factor of 1.35 for the presence of He, and the inferred SFR and $\tau_\mathrm{dep}$ for NGC\,300 digitized from \citet{2019Natur.569..519K} as triangles.
\label{fig:t_dep}}
\end{figure}

Figure \ref{fig:t_dep}, however, shows that the depletion time of molecular hydrogen gas is expected to change by nearly three orders of magnitude from $0.01Z_\odot$ to $Z_\odot$ in our simulations, while the SFR is changing only by a factor of ten over the same metallicity range. Thus, according to our simulations the star formation rate at low metallicities is a nonlinear function of the molecular mass, which implies that the fraction of stars forming in atomic gas increases with decreasing metallicity. 

As discussed by \citet{2012ApJ...759....9K}, the use of chemical equilibrium models to estimate H$_2$ abundance for star formation rate calculations can partly compensate for this trend and will result in higher SFR at low metallicity. However, this is hardly justified because this effectively trades one error for another by artificially overestimating $f_\mathrm{H_2}$ in the environment where the fraction is actually low. 
A much better alternative is to switch to a star formation prescription that captures the metallicity dependence of the actual star formation. 

\subsection{Comparisons with other models}

Motivated by the paucity of reliable \textsc{HI}-\textsc{H$_2$} transition models for the $Z \lesssim 0.2 \, Z_\odot$ regime (corresponding to metallicities typical of dwarf galaxies and galaxies at high redshift), we constructed fitting formulae for the molecular fraction as a function of gas density (or column density), local UV flux, and metallicity down to $Z = 0.01 Z_\odot$. This metallicity range includes the smallest metallicities observed in galaxies in the ultra-faint regime \citep{2017AA...606A.115H,Simon.2019}. Simultaneously, as we showed above these fitting formulae perform well at higher $Z$ up to the solar metallicity.

\begin{figure*}[ht!]
\epsscale{1.1}
\plotone{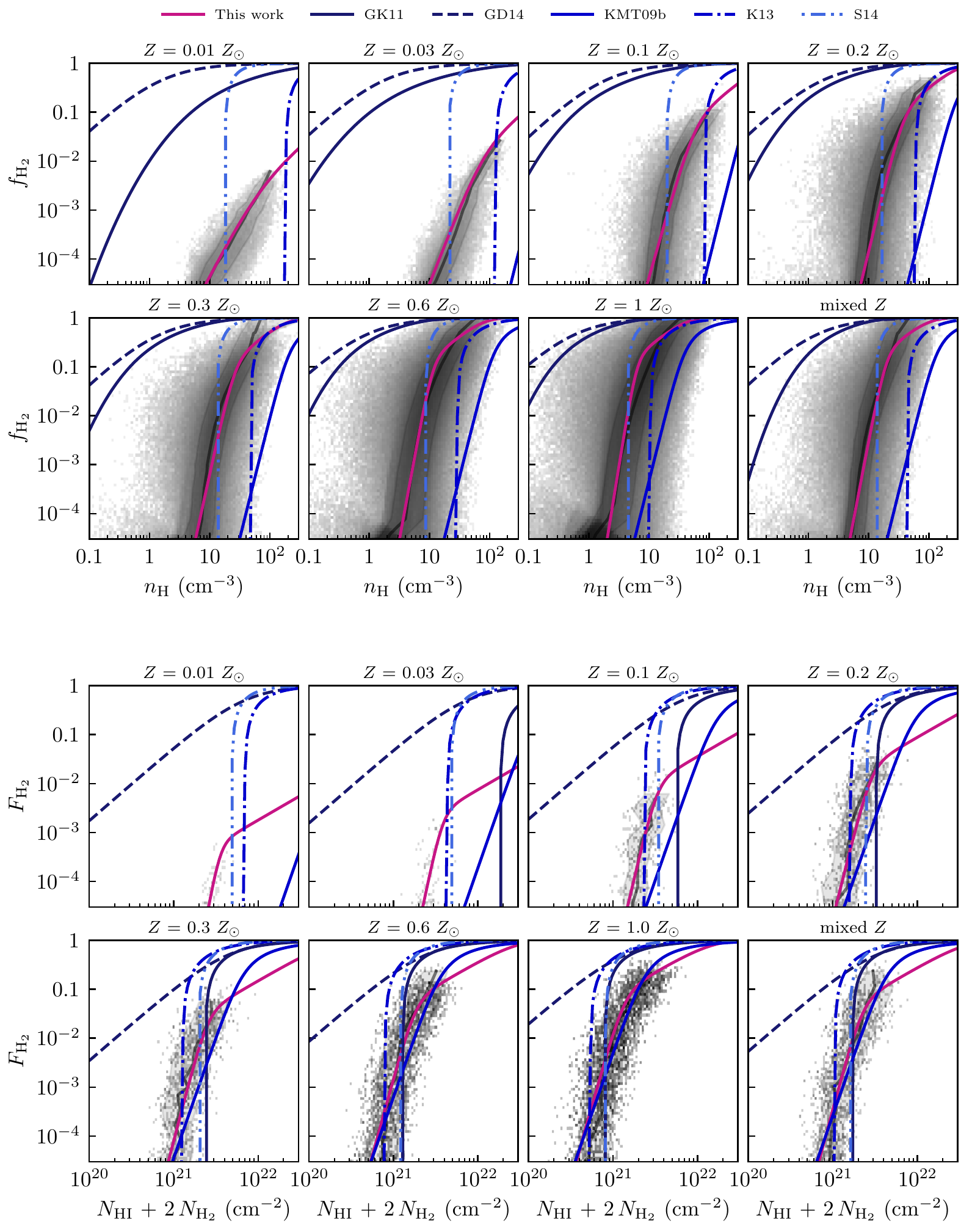}
\caption{The molecular hydrogen fraction as a function of $n_\mathrm{H}$ (\emph{top}) and $N_\mathrm{H}$ ($S = 100$ pc, \emph{bottom}) for different models including our own overplotted on the underlying $f_\mathrm{H_2}$ vs. $n_\mathrm{H}$ distribution from the simulation. For each model, we use the median $U_\mathrm{MW}$ ($U_\mathrm{MW} \sim 0.1$) and $S$ ($S \sim 10$ pc, \emph{top}; $S = 100$ pc, \emph{bottom}) for the high molecular fraction ($f_\mathrm{H_2} \ge 10^{-5}$) cells in the simulation. For our mixed metallicity run, we also use the median $Z$ ($\sim 0.5 \, Z_\odot$). For \krumholz\, and \sternberg, we assume a clumping factor of 1, and for \krumholz\, we also set $\rho_\mathrm{SD}$ (the density of stars and dark matter) to $\sim 0.1\; M_\odot$ pc$^{-3}$, which is the median for the high molecular fraction cells in the simulation with a non-zero $\rho_\mathrm{SD}$. We underplot the median density in bins of $f_\mathrm{H_2}$ in black for each run of the simulation.
\label{fig:fh2_models}}
\end{figure*}

Here we compare the fitting formulae presented in this paper and a number of models of molecular fraction in the literature \citep[namely,][]{2009ApJ...699..850K,2011ApJ...728...88G,2013MNRAS.436.2747K, 2014ApJ...790...10S, 2014ApJ...793...29G} with simulation results. The models parameterize average $f_{\rm H_2}$ and $F_{\rm H_2}$ as a function of gas density, metallicity, and UV field strength (the \kmt~model only accounts for the dependence on column density and metallicity; see Appendix \ref{app:model_implement} for details of the model implementations).

\begin{figure*}[ht!]
\epsscale{0.9}
\plotone{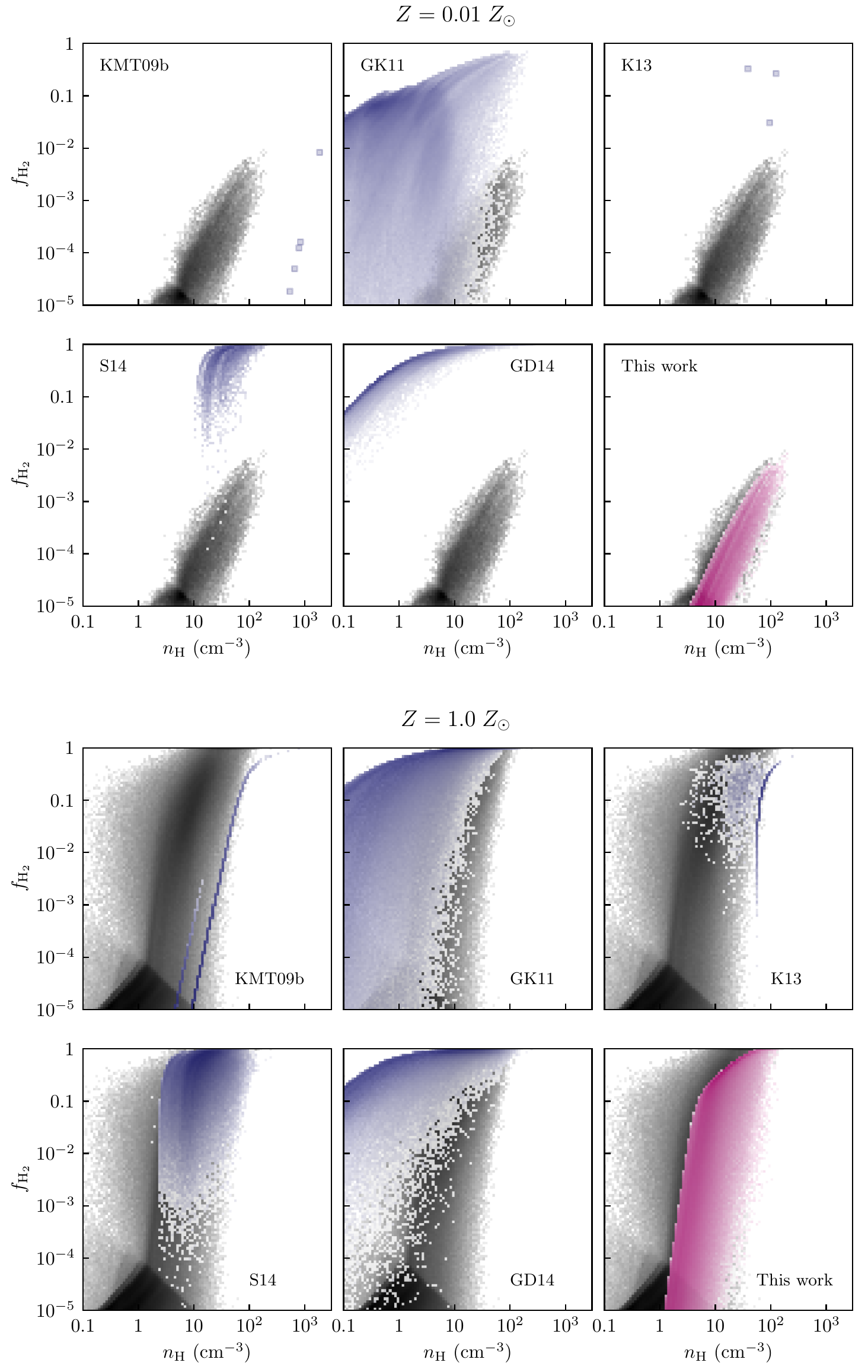}
\caption{Comparison of existing models and the \textsc{HI-H$_2$} model we present here on a volumetric cell-by-cell basis for $Z = 0.01 \, Z_\odot$ and $Z = 1.0 \, Z_\odot$. We use simulation values of density, metallicity, UV field strength, and grid cell size, and assume $f_c = 1$ for \krumholz\, and \sternberg\;as in Figure \ref{fig:fh2_models} and Table \ref{tab:acc}. Given how few \kmt\, and \krumholz\, model cells have $f_\mathrm{H_2} \ge 10^{-5}$ at densities consistent with those in the $Z = 0.01\, Z_\odot$ simulation, we plot the phase space location of each cell individually for these two models.
\label{fig:model_compare}}
\end{figure*}

\begin{figure*}[ht!]
\epsscale{0.9}
\plotone{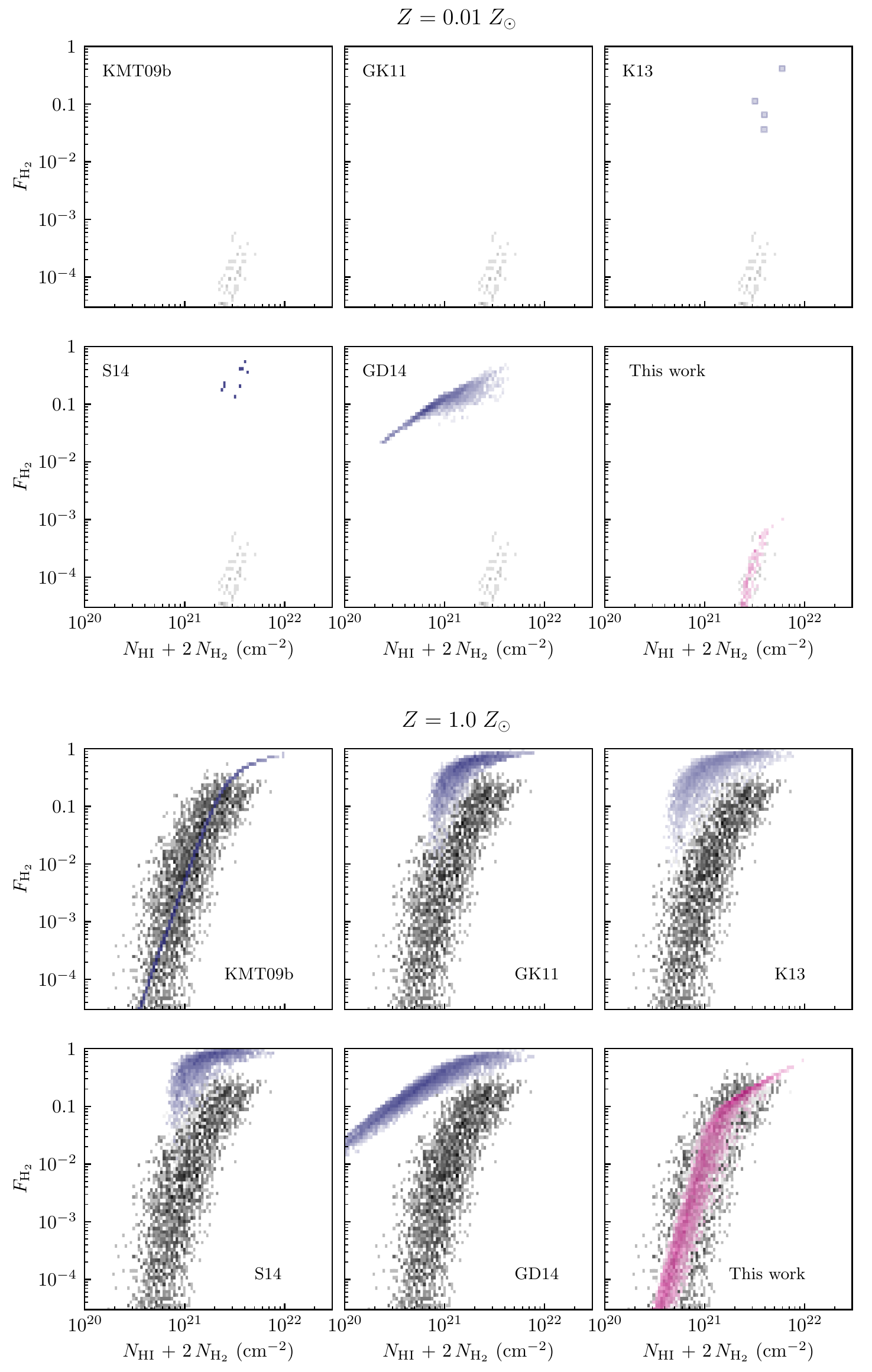}
\caption{Comparison of existing models and the \textsc{HI-H$_2$} model we present here on a projected cell-by-cell basis for $Z = 0.01 \, Z_\odot$ and $Z = 1.0 \, Z_\odot$. We use simulation values of density, metallicity, UV field strength, and scale, and assume $f_c = 1$ for \krumholz\, and \sternberg\;as in Figure \ref{fig:fh2_models} and Table \ref{tab:acc}. At low metallicities, neither \kmt\;nor \gk\; predict $f_\mathrm{H_2} (N_\mathrm{H})$ within the range shown here. As in Figure \ref{fig:model_compare}, given how few \krumholz\, model cells have $f_\mathrm{H_2} \ge 10^{-5}$ at densities consistent with those in the $Z = 0.01\, Z_\odot$ simulation, we plot the phase space location of each cell individually for that model. Unlike in Figure \ref{fig:model_compare}, where we compute $\rho_\mathrm{SD}$ on a per grid cell basis, we instead use a global value $\sim 0.1 \, M_\odot\, \mathrm{pc}^{-3}$ based on the median for each run.
\label{fig:projmodel_compare}}
\end{figure*}

Figure \ref{fig:fh2_models} shows comparisons of the average $f_{\rm H_2}$ and $F_{\rm H_2}$ to the molecular fraction as a function of volumetric and projected densities in simulations. In order to have a more direct comparison between existing models and the fits presented in this study, we also show predictions of each model when applied to the individual computational grid cells in our simulation for our single metallicity runs with the lowest and highest $Z$ in Figures \ref{fig:model_compare} and \ref{fig:projmodel_compare}. The very narrow distribution of $f_\mathrm{H_2}$ as a function of density in the \kmt\, model is likely due to its neglect of dependence on $U_\mathrm{MW}$, which drives the scatter in the simulation and in other models, including the model calibrated in this study.

Given that most existing models have been calibrated in the high-metallicity regime, it is not surprising that their accuracy improves with increasing metallicity (see Table \ref{tab:acc}). Interestingly, the accuracy of the \krumholz\, model is better at low metallicities for volumetric grid cells, but the match to the shape and behavior of the \textsc{HI-H$_2$} transition is better at high $Z$. This seems to be driven by their very steep functional form of $f_\mathrm{H_2}$, so that only a few grid cells with high $f_\mathrm{H_2}$ contribute to the inferred mass at lower $Z$, while at higher metallicity, the transition  $n_\mathrm{H}$ is underestimated, leading to the model underprediction of $M_\mathrm{H_2}$. 

In the case of the projected models, \kmt\, performs consistently well for $Z \gtrsim 0.1 \, Z_\odot$. The shape of $F_\mathrm{H_2}(N_\mathrm{H})$ is very close to what we observe in the simulation for these metallicities and the transition between atomic and molecular gas is consistent with the location of the transition in the simulation for $Z \gtrsim 0.4 \, Z_\odot$. \gk\, is the next most accurate of the existing models. The lowest metallicity runs result in an inferred $M_\mathrm{H_2} = 0 \, M_\odot$ with this model, but this is expected behavior for the projected \gk\, model, which is known to not be accurate for $Z \lesssim 0.01 \, Z_\odot$ \citep{2011ApJ...728...88G}.

With regard to the \gd\, model, it is worth noting that this model includes a phenomenological account for the H$_2$ self-shielding due to line overlap. 
This results in near independence of $f_{\rm H_2}$ on dust abundance (and thus metallicity) for $D\lesssim 0.2D_{\rm MW}$. The simulations used here do not include any accounting for such line overlap and thus a part of the difference of \gd\, model and simulation results at $Z\lesssim 0.2Z_\odot$ may be due to this difference. Otherwise, the \gk\, and \gd\, models are quite similar and thus the accuracy of the \gk\, model should be comparable to predictions of the \gd\, model without the line overlap effect. The similarity of these two models is the reason why their estimated molecular mass is similar at $Z\lesssim 0.2Z_\odot$ (see Table~\ref{tab:acc}). 

The fit presented in this paper reproduce  the total molecular hydrogen mass in the simulations considerably better than previous models.  Even restricting \krumholz\, to $Z \lesssim 0.1 \, Z_\odot$, where their model results in masses within a factor of $\sim$two of those measured in simulation, it appears to be a coincidence given the steep relation and few high $f_\mathrm{H_2}$ cells. 

\begin{deluxetable*}{clcccccccc}
\tablecaption{The ratio of H$_2$ mass in the simulation to model predicted value, $M_\mathrm{H_2,sim}$/$M_\mathrm{H_2,mod}$, for different models (both volumetric and projected), where masses are obtained by summing the actual and predicted H$_2$ value for every grid cell. We assume the same model parameters as in Figures \ref{fig:model_compare} and \ref{fig:projmodel_compare} respectively. For the projected models, we adopt show the accuracy for the representative $S = 100$ pc case. We denote model and metallicity combinations that produce no molecular hydrogen with a dash. \label{tab:acc}}
\tablehead{
\colhead{} & \colhead{Model} & \colhead{$0.01 \, Z_\odot$} & \colhead{$0.03 \, Z_\odot$} & \colhead{$0.1 \, Z_\odot$} & \colhead{$0.2 \, Z_\odot$} & \colhead{$0.3 \, Z_\odot$} & \colhead{$0.6 \, Z_\odot$} & \colhead{$1\, Z_\odot$} & \colhead{mixed $Z$}}
\startdata
\multirow{6}{*}{\rotatebox[origin=c]{90}{volumetric}}
& \kmt & 35.5 & $4.25 \times 10^3$ & 262 & 72.3 & 27.2 & 27.5 & 29.8 & 17.3\\
& \gk & $2.27 \times 10^{-4}$ & $2.38 \times 10^{-4}$ & $2.72 \times 10^{-3}$ & $1.30\times 10^{-2}$ & $2.27\times 10^{-2}$ & $5.27\times 10^{-2}$ & $8.83\times 10^{-2}$ & $5.72\times 10^{-2}$ \\
& \krumholz & 1.18 & 0.51 & 1.76 & 4.55 & 7.56 & 15.8 & 20.7 & 13.12\\
& \sternberg & $2.29\times 10^{-3}$& $6.20\times10^{-3}$& $3.39\times10^{-2}$ & 0.10 & 0.18 & 0.39 & 0.60 & 0.45\\
& \gd & $5.41 \times 10^{-5}$ & $1.91 \times 10^{-4}$ & $2.31 \times 10^{-3}$ & $1.09 \times 10^{-2}$ & $1.94 \times 10^{-2}$ & $4.50 \times 10^{-2}$ & $7.48 \times 10^{-2}$ & $5.06 \times 10^{-2}$\\
& This work & 1.01 & 0.89 & 1.12 & 1.26 & 1.20 & 1.02 & 0.75 & 0.79\\
\hline
\multirow{6}{*}{\rotatebox[origin=c]{90}{projected, 100 pc}}
& \kmt & $1.94\times10^4$ & $335$ & 2.37 & 1.49 & 0.82 & 0.57 & 0.48 & 0.40\\
& \gk & - & - & 1.42 & 0.25 & 0.18 & 0.16 & 0.16 & 0.16 \\
& \krumholz & $7.43 \times 10^{-2}$ & $1.88\times10^{-2}$ & $2.01\times10^{-2}$ & $3.82\times10^{-2}$ & $5.48\times10^{-2}$ & $9.70\times10^{-2}$ & 0.13 & 0.10\\
& \sternberg & $3.90\times 10^{-3}$& $1.53\times10^{-2}$& $4.17\times10^{-2}$ & $6.69\times10^{-2}$ & $8.00\times10^{-2}$ & 0.11 & 0.14 & 0.12\\
& \gd & $1.35 \times 10^{-4}$ & $5.98 \times 10^{-4}$ & $8.06 \times 10^{-3}$ & $3.46 \times 10^{-2}$ & $5.72 \times 10^{-2}$ & 0.1 & 0.13 & 0.11\\
& This work & 0.94 & 0.68 & 0.76 & 0.96 & 0.93 & 0.88 & 0.69 & 0.71\\
\enddata
\end{deluxetable*}

\section{Conclusions}
\label{sec:conclusions}

In this work we presented fitting formulae for volumetric (equations \ref{eq:fh2_vol}-\ref{eq:cD}) and projected (equations \ref{eq:fh2_proj}-\ref{eq:ncorr}) molecular gas fractions. The fits consist of a set of simple scalings calibrated to reproduce mean trends measured in simulations of a realistic dwarf galaxy similar to NGC 300 \citep{2021ApJ...918...13S}. Both volumetric and projected fits parameterize the molecular fraction as a function of gas density, gas metallicity, and the strength of the local free space ionizing UV field. 

Our main results and conclusions are as follows:
\begin{itemize}
    \item We show that the interstellar medium in the simulated galaxy changes qualitatively when gas metallicity is varied by two orders of magnitude. The density distribution becomes increasingly non-uniform as metallicity increases from $0.01Z_\odot$ to $Z_\odot$ (see Figure \ref{fig:disk}).
    \item In addition to the well-known trend of the \textsc{HI-H$_2$} transition shifting to higher densities with decreasing metallicity, the maximum achieved molecular fraction in the interstellar medium drops drastically to values much less than 1 at $Z \lesssim 0.2 \, Z_\odot$ (see Figures \ref{fig:fh2_vol} and \ref{fig:fh2_proj}), while the dependent of molecular fraction on density becomes less steep.
    \item We show that accurate fitting functions for volumetric and projected molecular fractions can be constructed if they account for the dependence on gas density, gas metallicity, the strength of the ionizing UV field (Figures \ref{fig:fit_compare} and \ref{fig:nmed}). We demonstrate that the presented fits reproduce the dependence of the \textsc{HI-H$_2$} transition on metallicity and the overall shape of the molecular fraction-density distribution than existing models (Figures~\ref{fig:fh2_models}--\ref{fig:projmodel_compare})
    
    \item We also show that the volumetric (projected) molecular fraction fit is applied to individual cells (projected patches) reproduces the total molecular mass in the simulated galaxies to a factor $\lesssim 1.25$ ($\lesssim 1.5$)  across the entire explored range of galaxy metallicities (Figures \ref{fig:acc} and \ref{fig:time_compare}). This is considerably better than the estimates using existing models of the molecular hydrogen fraction and  \textsc{HI-H$_2$} transition (see Figures \ref{fig:fh2_models} and \ref{fig:model_compare} and Table \ref{tab:acc}).
\end{itemize}

The presented model should be useful in modeling molecular gas abundance in simulations that do not include explicit modeling of H$_2$ and low-resolution simulations and semi-analytical models. However, we argue that star formation modeling in simulations should not be based on molecular gas fraction because our simulation results indicate that star formation rate becomes a non-linear function of molecular gas density at metallicities $<0.1Z_\odot$ due to non-equilibrium effects (see Section~\ref{sec:discussion} and Fig.~\ref{fig:t_dep}). As an alternative, the star formation efficiency and depletion time can be calibrated using such simulations and we will present such calibrations in the follow-up work.  

\begin{acknowledgments}
The simulations used in this work were carried out on the Midway cluster maintained by the University of Chicago Research Computing Center. A.K. was supported by the National Science Foundation grants AST-1714658 and AST-1911111 and NASA ATP grant 80NSSC20K0512.
V.S. is grateful for the support provided by Harvard University through the Institute for Theory and Computation Fellowship.
\end{acknowledgments}

\software{Astropy \citep{astropy:2013, astropy:2018, astropy:2022}, Matplotlib \citep{Hunter:2007}, NumPy \citep{harris2020array}, SciPy \citep{2020SciPy-NMeth}, yt \citep{2011ApJS..192....9T}, WebPlotDigitizer \citep{Rohatgi2022}}

\appendix
\section{Implementation of other HI-H$_2$ models}
\label{app:model_implement}

Given that there are different formulations  of the HI-H$_2$ transition models which we use in comparisons in Section~\ref{sec:discussion}, we offer details of how $f_\mathrm{H_2}$ and $F_{\rm H_2}$ were computed as a function of density and column density for each model.

\subsection{\gk\;and \gd}
\label{subsec:gg}
For both \gk\;and \gd models, we follow the model as presented in the original papers \citep{2011ApJ...728...88G, 2014ApJ...795...37G}. We set the scale in \gd\; to be the size of the simulation's grid cells.

For the projected case of \gk, we use the high density approximation
\begin{equation}
 \Sigma_\mathrm{HI} = 2\, \Sigma_c = 40\, \mathrm{M_\odot\, pc^{-2}} \frac{\Lambda^{4/7}}{D_\mathrm{MW}} \frac{1}{\sqrt{1 + U_\mathrm{MW}D_\mathrm{MW}^2}}   
\end{equation}
to compute 
\begin{equation}
    f_\mathrm{H_2} = 1 - \frac{\Sigma_\mathrm{HI}}{\Sigma_\mathrm{H}}
\end{equation}
Here we assume that $\Sigma_\mathrm{H} = \Sigma_\mathrm{H_2} + \Sigma_\mathrm{HI}$, consistent with a nonexistent or negligible amount of ionization. If the expression produces $\Sigma_\mathrm{HI} > \Sigma_\mathrm{H}$, we set $f_\mathrm{H_2} = 0$. We note that assuming $\Sigma_\mathrm{HI} \simeq \Sigma_c$ results in somewhat better agreement between \gk\;and the location of the HI-H$_2$ transition in the simulation, particularly at low $Z$ , though we do not show it here, opting instead to use the formulation from the paper.

In the case of \gd, we follow their Equations 8 - 10 directly and compute molecular fraction as $f_\mathrm{H_2} = R_\mathrm{mol}/(1 + R_\mathrm{mol})$.

\subsection{The \kmt\;and \krumholz~models}
\label{subsec:kk}
We use the \kmt\;model \citet{2009ApJ...699..850K} rather than the model presented in \citet{2009ApJ...693..216K} due to its better accuracy at low molecular fractions, consistent with the low $Z$ regime.
The \kmt\; model uses surface density of the atomic-molecular complex $\Sigma_\mathrm{comp} = c\Sigma_\mathrm{gas}$ with the clumping factor $c \to 1$ on scales $\sim100$ pc. We approximate $\Sigma_\mathrm{comp} \approx \Sigma_\mathrm{H}$, since the grid cells in the simulation with $f_\mathrm{H_2} \ge 10^{-5}$ are generally smaller than 100 pc. We note that this is not a perfect approximation since this expression for $\Sigma_\mathrm{comp}$ should be averaged on $\sim 100$ pc scales. In the volumetric case, to compute $\Sigma_\mathrm{H}$ from $n_\mathrm{H}$, we take $\Sigma_\mathrm{H} = (n_\mathrm{H}/\mathrm{cm^{-3}})\, (L/\mathrm{cm})\,(m_p/\mathrm{g})$, which we then convert to units of $M_\odot\, \mathrm{pc}^{-2}$. Similarly in the \krumholz\;model, we compute $\Sigma_0$ in the same way as $\Sigma_\mathrm{H}$.

We largely follow \citet{2013MNRAS.436.2747K} for the implementation of the \krumholz~model. In order to compute $f_\mathrm{H_2}$, we first need to compute $n_\mathrm{CNM, hydro}$, which relies on $P_\mathrm{th}$. $P_\mathrm{th}$ is a function of $R_\mathrm{mol}$ $ = f_\mathrm{H_2}/(1 - f_\mathrm{H_2})$. We ultimately iterate over this step 10 times, following the more efficient approach in \citet{2018ApJS..238...33D}, where $f_\mathrm{H_2}$ is initialized at 0.5 and is averaged as it advances through the iteration so that for step $i$, $f_\mathrm{H_2, i} \approx 0.3 f_\mathrm{H_2, i - 1} + 0.7 f_\mathrm{H_2, i}$.

We also adopt a clumping factor, $f_c = 1$. For Figure \ref{fig:fh2_models} we adopt the  median density $\rho_\mathrm{SD}=0.1 \;M_\odot \,\mathrm{pc^{-2}}$ in our simulation in regions where $f_\mathrm{H_2} \ge 10^{-5}$ and $\rho_\mathrm{SD}$ is non-zero. As in \citet[\S C.4]{2018ApJS..238...33D}, we find no significant difference between using different constant values of $\rho_\mathrm{SD}$, but that using different values on a cell-by-cell basis (as we do in Figure \ref{fig:model_compare}, where $\rho_\mathrm{SD}$ is computed directly from the simulation) has a significant effect.

\subsection{The \sternberg~model}

We follow the simple \citet[\S C.5]{2018ApJS..238...33D} formulation for the \sternberg\; model \citep{2014ApJ...790...10S, 2016ApJ...822...83B}. To convert the number density to column density, we assume $N_\mathrm{H}/\mathrm{cm^{-2}} = (n_\mathrm{H}/\mathrm{cm^{-3}})(L/\mathrm{cm})$ for comparison with our volumetric model. In the same vein, when comparing with our projected model, we take $n_\mathrm{H}/\mathrm{cm}^{-3} = (N_\mathrm{H}/\mathrm{cm}^{-2})(1.54\times10^{21}/\mathrm{cm})^{-1}$, which corresponds to the approximate height of the disk ($\sim 500$ pc).

\bibliography{references,vs}{}
\bibliographystyle{aasjournal}

\end{document}